\documentclass[aps,prl,showpacs,twocolumn,groupedaddress]{revtex4}

\usepackage{graphicx}
\usepackage{dcolumn}
\usepackage{bm}
\usepackage{amssymb}

\def\met{\mbox{${\hbox{$E$\kern-0.6em\lower-.1ex\hbox{/}}}_T$}}

\begin{document}

\hspace{5.2in} \mbox{FERMILAB-PUB-05/061-E}

\title{Production of {\it WZ} Events in $p\bar{p}$ Collisions
 at $\sqrt{s}=1.96$ TeV and Limits on Anomalous {\it WWZ} Couplings}

\date{April 11, 2005}
%
\author{                                                                      
V.M.~Abazov,$^{35}$                                                           
B.~Abbott,$^{72}$                                                             
M.~Abolins,$^{63}$                                                            
B.S.~Acharya,$^{29}$                                                          
M.~Adams,$^{50}$                                                              
T.~Adams,$^{48}$                                                              
M.~Agelou,$^{18}$                                                             
J.-L.~Agram,$^{19}$                                                           
S.H.~Ahn,$^{31}$                                                              
M.~Ahsan,$^{57}$
% ...
\\                                                                            
\vskip 0.30cm                                                                 
\centerline{(D\O\ Collaboration)}                                             
\vskip 0.30cm                                                                 
}                                                                             
\affiliation{                                                                 
\centerline{$^{1}$Universidad de Buenos Aires, Buenos Aires, Argentina}       
\centerline{$^{2}$LAFEX, Centro Brasileiro de Pesquisas F{\'\i}sicas,         
                  Rio de Janeiro, Brazil}                                     
\centerline{$^{3}$Universidade do Estado do Rio de Janeiro,                   
                  Rio de Janeiro, Brazil}                                     
\centerline{$^{4}$Instituto de F\'{\i}sica Te\'orica, Universidade            
                  Estadual Paulista, S\~ao Paulo, Brazil}                     
\centerline{$^{5}$University of Alberta, Edmonton, Alberta, Canada,           
               Simon Fraser University, Burnaby, British Columbia, Canada,}   
\centerline{York University, Toronto, Ontario, Canada, and                    
         McGill University, Montreal, Quebec, Canada}                         
\centerline{$^{6}$Institute of High Energy Physics, Beijing,                  
                  People's Republic of China}                                 
\centerline{$^{7}$University of Science and Technology of China, Hefei,       
                  People's Republic of China}                                 
\centerline{$^{8}$Universidad de los Andes, Bogot\'{a}, Colombia}             
\centerline{$^{9}$Center for Particle Physics, Charles University,            
                  Prague, Czech Republic}                                     
\centerline{$^{10}$Czech Technical University, Prague, Czech Republic}        
\centerline{$^{11}$Institute of Physics, Academy of Sciences, Center          
                  for Particle Physics, Prague, Czech Republic}               
\centerline{$^{12}$Universidad San Francisco de Quito, Quito, Ecuador}        
\centerline{$^{13}$Laboratoire de Physique Corpusculaire, IN2P3-CNRS,         
                 Universit\'e Blaise Pascal, Clermont-Ferrand, France}        
\centerline{$^{14}$Laboratoire de Physique Subatomique et de Cosmologie,      
                  IN2P3-CNRS, Universite de Grenoble 1, Grenoble, France}     
\centerline{$^{15}$CPPM, IN2P3-CNRS, Universit\'e de la M\'editerran\'ee,     
                  Marseille, France}                                          
\centerline{$^{16}$Laboratoire de l'Acc\'el\'erateur Lin\'eaire,              
                  IN2P3-CNRS, Orsay, France}                                  
\centerline{$^{17}$LPNHE, IN2P3-CNRS, Universit\'es Paris VI and VII,         
                  Paris, France}                                              
\centerline{$^{18}$DAPNIA/Service de Physique des Particules, CEA, Saclay,    
                  France}                                                     
\centerline{$^{19}$IReS, IN2P3-CNRS, Universit\'e Louis Pasteur, Strasbourg,  
                France, and Universit\'e de Haute Alsace, Mulhouse, France}   
\centerline{$^{20}$Institut de Physique Nucl\'eaire de Lyon, IN2P3-CNRS,      
                   Universit\'e Claude Bernard, Villeurbanne, France}         
\centerline{$^{21}$III. Physikalisches Institut A, RWTH Aachen,               
                   Aachen, Germany}                                           
\centerline{$^{22}$Physikalisches Institut, Universit{\"a}t Bonn,             
                  Bonn, Germany}                                              
\centerline{$^{23}$Physikalisches Institut, Universit{\"a}t Freiburg,         
                  Freiburg, Germany}                                          
\centerline{$^{24}$Institut f{\"u}r Physik, Universit{\"a}t Mainz,            
                  Mainz, Germany}                                             
\centerline{$^{25}$Ludwig-Maximilians-Universit{\"a}t M{\"u}nchen,            
                   M{\"u}nchen, Germany}                                      
\centerline{$^{26}$Fachbereich Physik, University of Wuppertal,               
                   Wuppertal, Germany}                                        
\centerline{$^{27}$Panjab University, Chandigarh, India}                      
\centerline{$^{28}$Delhi University, Delhi, India}                            
\centerline{$^{29}$Tata Institute of Fundamental Research, Mumbai, India}     
\centerline{$^{30}$University College Dublin, Dublin, Ireland}                
\centerline{$^{31}$Korea Detector Laboratory, Korea University,               
                   Seoul, Korea}                                              
\centerline{$^{32}$CINVESTAV, Mexico City, Mexico}                            
\centerline{$^{33}$FOM-Institute NIKHEF and University of                     
                  Amsterdam/NIKHEF, Amsterdam, The Netherlands}               
\centerline{$^{34}$Radboud University Nijmegen/NIKHEF, Nijmegen, The          
                  Netherlands}                                                
\centerline{$^{35}$Joint Institute for Nuclear Research, Dubna, Russia}       
\centerline{$^{36}$Institute for Theoretical and Experimental Physics,        
                  Moscow, Russia}                                             
\centerline{$^{37}$Moscow State University, Moscow, Russia}                   
\centerline{$^{38}$Institute for High Energy Physics, Protvino, Russia}       
\centerline{$^{39}$Petersburg Nuclear Physics Institute,                      
                   St. Petersburg, Russia}                                    
\centerline{$^{40}$Lund University, Lund, Sweden, Royal Institute of          
                   Technology and Stockholm University, Stockholm,            
                   Sweden, and}                                               
\centerline{Uppsala University, Uppsala, Sweden}                              
\centerline{$^{41}$Lancaster University, Lancaster, United Kingdom}           
\centerline{$^{42}$Imperial College, London, United Kingdom}                  
\centerline{$^{43}$University of Manchester, Manchester, United Kingdom}      
\centerline{$^{44}$University of Arizona, Tucson, Arizona 85721, USA}         
\centerline{$^{45}$Lawrence Berkeley National Laboratory and University of    
                  California, Berkeley, California 94720, USA}                
\centerline{$^{46}$California State University, Fresno, California 93740, USA}
\centerline{$^{47}$University of California, Riverside, California 92521, USA}
\centerline{$^{48}$Florida State University, Tallahassee, Florida 32306, USA} 
\centerline{$^{49}$Fermi National Accelerator Laboratory, Batavia,            
                   Illinois 60510, USA}                                       
\centerline{$^{50}$University of Illinois at Chicago, Chicago,                
                   Illinois 60607, USA}                                       
\centerline{$^{51}$Northern Illinois University, DeKalb, Illinois 60115, USA} 
\centerline{$^{52}$Northwestern University, Evanston, Illinois 60208, USA}    
\centerline{$^{53}$Indiana University, Bloomington, Indiana 47405, USA}       
\centerline{$^{54}$University of Notre Dame, Notre Dame, Indiana 46556, USA}  
\centerline{$^{55}$Iowa State University, Ames, Iowa 50011, USA}              
\centerline{$^{56}$University of Kansas, Lawrence, Kansas 66045, USA}         
\centerline{$^{57}$Kansas State University, Manhattan, Kansas 66506, USA}     
\centerline{$^{58}$Louisiana Tech University, Ruston, Louisiana 71272, USA}   
\centerline{$^{59}$University of Maryland, College Park, Maryland 20742, USA} 
\centerline{$^{60}$Boston University, Boston, Massachusetts 02215, USA}       
\centerline{$^{61}$Northeastern University, Boston, Massachusetts 02115, USA} 
\centerline{$^{62}$University of Michigan, Ann Arbor, Michigan 48109, USA}    
\centerline{$^{63}$Michigan State University, East Lansing, Michigan 48824,   
                   USA}                                                       
\centerline{$^{64}$University of Mississippi, University, Mississippi 38677,  
                   USA}                                                       
\centerline{$^{65}$University of Nebraska, Lincoln, Nebraska 68588, USA}      
\centerline{$^{66}$Princeton University, Princeton, New Jersey 08544, USA}    
\centerline{$^{67}$Columbia University, New York, New York 10027, USA}        
\centerline{$^{68}$University of Rochester, Rochester, New York 14627, USA}   
\centerline{$^{69}$State University of New York, Stony Brook,                 
                   New York 11794, USA}                                       
\centerline{$^{70}$Brookhaven National Laboratory, Upton, New York 11973, USA}
\centerline{$^{71}$Langston University, Langston, Oklahoma 73050, USA}        
\centerline{$^{72}$University of Oklahoma, Norman, Oklahoma 73019, USA}       
\centerline{$^{73}$Brown University, Providence, Rhode Island 02912, USA}     
\centerline{$^{74}$University of Texas, Arlington, Texas 76019, USA}          
\centerline{$^{75}$Southern Methodist University, Dallas, Texas 75275, USA}   
\centerline{$^{76}$Rice University, Houston, Texas 77005, USA}                
\centerline{$^{77}$University of Virginia, Charlottesville, Virginia 22901,   
                   USA}                                                       
\centerline{$^{78}$University of Washington, Seattle, Washington 98195, USA}  
}                                                                             

\begin{abstract}
We present results from a search for {\it WZ} production with 
subsequent decay to 
$\ell\nu \ell'\bar{\ell'}$ $(\ell$ and $\ell'=e$ or $\mu)$ using  
$0.30$ fb$^{-1}$ of data collected by the D\O\ experiment between
2002 and 2004 at the Tevatron. 
Three events with {\it WZ} decay characteristics are observed.
With an estimated background of $0.71\pm0.08$ events, 
we measure the {\it WZ} production cross 
section to be $4.5_{-2.6}^{+3.8}$~\textrm{pb}, with 
a 95\% C.L. upper limit of 
$13.3$~\textrm{pb}.  
The 95\% C.L. limits for anomalous {\it WWZ} couplings
are found to be $-2.0<\Delta\kappa _{Z}<2.4$ for form factor scale 
$\Lambda$ = 1 TeV, and $-0.48<\lambda _{Z}<0.48$ and 
$-0.49<\Delta g_{1}^{Z}<0.66$ for $\Lambda$ = 1.5 TeV.
\end{abstract}

\author{}
\maketitle

\begin{figure}[tbh]
\hskip 0cm \includegraphics[width=3.25in]{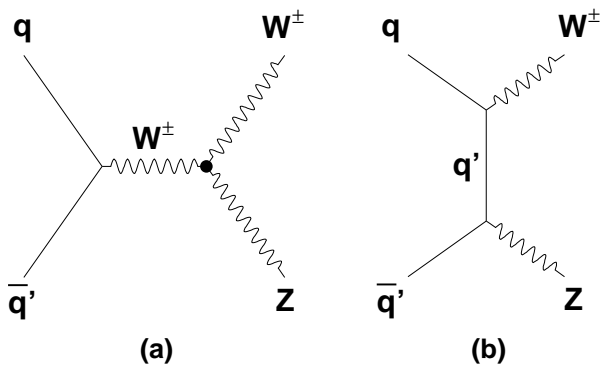}
\caption{Tree-level diagrams for {\it WZ} production in $p\bar{p}$
 collisions. Diagram (a) contains the {\it WWZ} trilinear gauge 
 coupling vertex.}
\label{fig:diag}
\end{figure}

The $SU(2)_{L}\otimes U(1)_{Y}$ structure of the standard model (SM) 
Lagrangian implies that the electroweak gauge bosons 
$W$ and $Z$ interact with one another through trilinear 
and quartic vertices. As a consequence,
the production cross section $\sigma(p\bar{p}\rightarrow{\it WZ})$ depends
on the {\it WWZ} gauge coupling shown in Fig.~\ref{fig:diag}a. 
The SM predicts that the strength of that coupling is $-e\cot{\theta_W}$, 
where $e$ is the electric charge and $\theta_W$ is the weak mixing angle.
More generally, excursions of the {\it WWZ} interactions from the SM can
be described by an effective Lagrangian with parameters $g_{1}^{Z}\ $, 
$\lambda_{Z}\ $ and $\kappa_{Z}$~\cite{ref:bysm}. This effective
Lagrangian reduces to the SM Lagrangian when the couplings are set to their 
SM values $g_{1}^{Z}=\kappa_{Z}=1$ and $ \lambda_{Z}=0$.  
Non-SM values of these couplings will increase $\sigma_{\it WZ}$.  
Therefore a measurement of the {\it WZ}
production cross section provides a sensitive test of the strength of 
the {\it WWZ} interaction. 
This test also probes for low-energy manifestations of new physics, 
appearing at a 
higher mass scale, that complements searches to be carried out with future 
higher-energy accelerators.

A model-independent test for anomalous trilinear boson couplings using 
$\sigma_{\it WZ}$ is unique among vector boson pair production processes 
in that {\it WZ} diagrams contain only {\it WWZ}, and 
not $WW\gamma$, vertices. 
Anomalous trilinear gauge boson coupling limits set using
characteristics of $W^+W^-$ 
production~\cite{CDF1, aleph, delphi, L3, opal, LEPass, run1wz}
are sensitive to both the $WW\gamma$ and {\it WWZ} couplings
and must make an assumption~\cite{LEPass, HISZ} relating them. 
Furthermore, as the $W^{\pm}Z$ production process is 
unavailable at $e^+ e^-$ colliders~\cite{aleph, delphi, L3, opal},
a hadron collider such as the Tevatron at Fermilab 
provides an unique opportunity 
for measurement of the {\it WWZ} coupling.

Using 90 pb$^{-1}$ of $p\bar{p}$ collisions collected at
$\sqrt{s}=1.8$ TeV during Run~I (1992--1996), the D\O \ Collaboration
established that 
$\sigma_{\it WZ}<47$~\textrm{pb} at 95\% C.L. From these data, D\O\ also set 
95\% C.L. limits $|g_{1}^{Z}-1|<1.63$ and $|\lambda_{Z}|<1.42$ for 
a form factor scale~\cite{ref:bysm} $\Lambda$ = 1 TeV~\cite{run1wz}. 
With a higher center-of-mass energy ($\sqrt{s}$ =1.96 \textrm{TeV}) 
expected to increase the SM {\it WZ} production cross section 
to 3.7$\pm$0.1 \textrm{pb}~\cite{ref:wzxsec}, more
luminosity, and improved detectors, the Run~II Tevatron program 
opens a new window for studies of {\it WZ} production. 
The CDF Collaboration recently announced
a $15.2$ pb upper limit at the 95\% C.L. on the combined cross section for 
{\it WZ} and {\it ZZ} production~\cite{ref:CDFZZWZ} .

We present the results of a search for {\it WZ} production with
``trilepton'' final states 
$\ell\nu \ell'\bar{\ell'}$ $(\ell$ and $\ell'=e$ or $\mu)$ 
using data collected by the D\O\ experiment from 2002--2004 
at $\sqrt{s}=1.96$ TeV. Requiring three isolated high
transverse momentum $(p_T)$ charged leptons and large missing
transverse energy
(\mbox{${\hbox{$E$\kern-0.6em\lower-.1ex\hbox{/}}}_T$}), 
to indicate the presence of a neutrino, strongly suppresses backgrounds
which mimic the {\it WZ} signal. However, branching ratios 
sum to only $1.5\%$ for trilepton final states 
($\mu\nu ee$, $e\nu\mu\mu$, $e\nu ee$ and 
$\mu\nu\mu\mu$).  The {\it WZ} signal that we seek is distinct but rare. 

The D\O\ detector~\cite{ref:run1det,ref:run2det} 
comprises several subdetectors 
and a trigger and data acquisition system. The central-tracking system
consists of a silicon microstrip tracker (SMT) and a central fiber tracker
(CFT) located within a 2~T superconducting solenoidal magnet. 
The SMT and CFT measure the locations of the collisions and the momenta 
of charged particles. The energies of electrons,
photons, and hadrons, and the amount of 
\mbox{${\hbox{$E$\kern-0.6em\lower-.1ex\hbox{/}}}_T$},
is measured in three uranium/liquid-argon 
calorimeters, each housed in a separate cryostat~\cite{ref:run1det}: 
a central section (CC) covering $|\eta|\leq 1.1$
and two end calorimeters (EC) extending coverage to $|\eta|\leq 4.2$,
where $\eta$ is the pseudorapidity. 
Scintillators between the CC and EC
cryostats provide sampling of developing showers for $1.1<|\eta|<1.4$. 
A muon system~\cite{ref:run2det} resides beyond the calorimetry, 
and consists of a layer of
tracking detectors and scintillation trigger counters in front of 
1.8~T toroidal magnets, followed by two similar layers behind the toroids. 
A three level trigger and data acquisition system uses information
from the subdetectors to select $\approx$ 50~Hz of collisions for 
further ``offline'' reconstruction.

With at least three high-$p_{T}$ charged leptons in the candidate
events, the overall trigger efficiency for the {\it WZ} signal 
is nearly 100\%.
Integrated luminosities for the 
$e\nu ee$, $\mu\nu ee$, $e\nu\mu\mu$ and $\mu\nu\mu\mu$ final
states are 320 $\mathrm{pb}^{-1}$, 290 $\mathrm{pb}^{-1}$, 
280 $\mathrm{pb}^{-1}$, 
and 290 $\mathrm{pb}^{-1}$, respectively, with a common uncertainty of 
$6.5\%$~\cite{lumy}. 

Electrons from $W$ and $Z$ boson decays are identified by their 
pattern of spatially isolated energy deposition in the calorimeter and by 
the presence of a matching track in the central tracking system.  
The transverse energy of an electron, measured in the calorimeter, 
must satisfy $E_{T}$ $>$ 15 \textrm{GeV}.

A muon is identified by a pattern of hits in the scintillation
counter and drift chamber system and must have a matching
central track. Muon isolation is determined from an
examination of the energy in calorimeter cells and the
momenta of any additional tracks around the muon.
Muons must have $p_{T}>15$~\textrm{\ GeV/$c$}.

Missing transverse energy
is determined from the negative of the vector sum of transverse
energies of the calorimeter cells, adjusted for the 
presence of any muons identified above. 

\begin{figure}[ptb]
\vspace{-0.5in}
\includegraphics[width=3.3in]{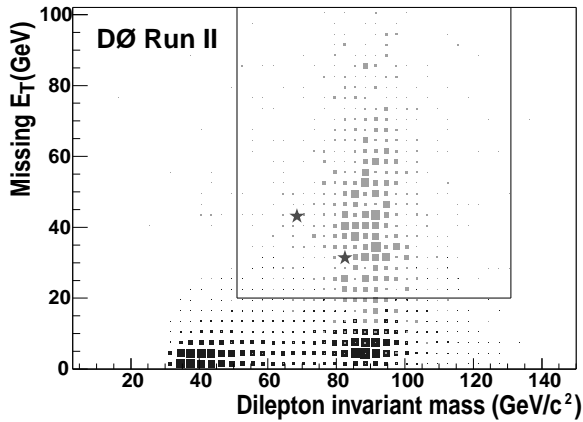}
\caption{\mbox{${\hbox{$E$\kern-0.6em\lower-.1ex%
\hbox{/}}}_T$}~versus dilepton invariant mass distribution for 
$\sim 200$ fb$^{-1}$ of simulated 
$WZ\rightarrow \mu\nu\mu\mu$ events (light grey)
and for expected $Z+{\rm jet(s)}$ background events (dark grey). 
The central box
shows the event selection criteria. The two $WZ\rightarrow \mu\nu\mu\mu$ 
candidates are indicated as stars. The corresponding figures 
are similar in the 
channels where the $Z$ boson decays to electrons.  
There is one candidate for the 
$WZ\rightarrow e\nu ee$ decay channel.}
\label{fig:massmet}
\end{figure}

The {\it WZ} event selection requires at least three charged leptons 
that originate from a common interaction vertex and survive
the electron or muon identification criteria outlined above. 
To associate reconstructed tracks with leptons unambiguously, they
are required to be spatially separated.
To select $Z$ bosons and suppress backgrounds further, the invariant mass of
a like-flavor lepton pair must fall within 
$71$ \textrm{GeV}/$c^2$ to 111 \textrm{GeV}/$c^2$ for 
$e^{+}e^{-}$ events, and 51 \textrm{GeV}/$c^2$ to 131 \textrm{GeV}/$c^2$ for 
$\mu^{+}\mu^{-}$ events, where the different mass windows correspond
to the respective resolutions of the calorimeter and the central tracker. 
For the $e\nu ee$ and $\mu\nu\mu\mu$ channels, the lepton pair with 
invariant mass
closest to the $Z$ boson mass is chosen as the $Z$ candidate. The 
\mbox{${\hbox{$E$\kern-0.6em\lower-.1ex\hbox{/}}}_T$} is required to 
be greater than $20$ GeV, consistent with a $W$ boson decay. 
The transverse mass, although not used as a selection criterion, 
is calculated from the $p_{T}$ of the unpaired third lepton
and the \mbox{${\hbox{$E$\kern-0.6em\lower-.1ex\hbox{/}}}_T$}.
Finally, to reject background from $t\bar{t}$ events, the vector sum of the
transverse energies in all calorimeter cells, excluding the leptons, must be
less than 50 \textrm{GeV}. 
Figure~\ref{fig:massmet} shows the comparison of 
the dilepton invariant mass and 
\mbox{${\hbox{$E$\kern-0.6em\lower-.1ex\hbox{/}}}_T$} 
distributions expected for $WZ\rightarrow \mu\nu\mu\mu$ events
to the background from $Z+{\rm jet(s)}$ events.

Applying all selection requirements leaves one $e\nu ee$ and two 
$\mu\nu\mu\mu$ candidates. Table~\ref{tab:wzcand} summarizes the 
kinematic properties of these events.

\begin{table*}[tbh]
\caption{Kinematic properties of the three {\it WZ} candidates. Provided 
are the momentum four-vectors for the two leptons which constitute the
$Z$ boson candidate, the invariant mass formed from those two leptons, 
the momentum 4-vector of the charged lepton from the 
$W$ boson decay, the components of 
the $\mbox{${\hbox{$E$\kern-0.6em\lower-.1ex\hbox{/}}}_T$}$,
and the transverse mass computed from the third lepton and the 
$\mbox{${\hbox{$E$\kern-0.6em\lower-.1ex\hbox{/}}}_T$}$~\cite{coords}.
The units are GeV, GeV/$c$, GeV/$c^2$, 
as appropriate.}
\begin{center}
\begin{ruledtabular}
\begin{tabular}{c|cccc|cccc|c|cccc|ccc} 
Final       & \multicolumn{4}{c|}{$\ell _{Z}$}
            & \multicolumn{4}{c|}{$\ell _{Z}$}
              &  
            & \multicolumn{4}{c|}{$\ell _{W}$}
              &
              & 
              &                     \\ 
State & $p_x$ &$p_y$  & $p_z$ & $E$ & 
        $p_x$ &$p_y$  & $p_z$ & $E$ & $m_{\ell \ell }$ 
              &      
        $p_x$ &$p_y$  & $p_z$ & $E$  
              &$\mbox{${\hbox{$E$\kern-0.6em\lower-.1ex\hbox{/}}}_T$}_x$
              &$\mbox{${\hbox{$E$\kern-0.6em\lower-.1ex\hbox{/}}}_T$}_y$
              & $m_{T}$             \\ \hline
$e\nu ee$ 
      & $-47.3$ & $-25.9$ & $292$  & $297$  & 
        $13.3$  & $37.6$  & $111$  & $118$  & 
        $91.9$  &  
        $45.3$  & $-32.1$ & $-16.5$ & $57.9$  & 
        $-19.6$  & $-23.5$  & 
                                        72.3 \\ \hline
$\mu \nu \mu \mu $
      &  24.5 &  11.6 & 29.7  & 40.2  & 
        $-38.7$ & $-12.4$ & $-17.1$ & 44.1  & 
         82.1 &  
        $-19.3$ & $-16.7$ & 101  & 105  &
        $24.1$ & 19.8 & 
                                        56.4 \\ \hline 
$\mu \nu \mu \mu $ 
      & $-15.1$ &  19.9 & 24.4  & 35.0  &
         20.2 & $-42.5$ & 57.1  & 74.0  & 
         68.5 &  
        $-21.9$ & $-5.90$ & $-16.4$ & 28.0  & 
         34.8 & 25.4  & 
                                        62.5 \\ 
\end{tabular}
\end{ruledtabular}
\end{center}
\label{tab:wzcand}
\end{table*}

Signal acceptances include geometric and kinematic effects and are 
obtained using
Monte Carlo samples produced with the {\sc PYTHIA} event generator 
\cite{ref:pythia} followed by the {\sc GEANT}-based~\cite{ref:geant} 
D\O\ detector-simulation program.  Acceptances are calculated by counting
the number of events that pass all selection criteria, except the lepton
identification and track-matching requirements. The results are 
$0.283\pm0.009$, $0.279\pm0.008$, $0.287\pm0.009$ and $0.294\pm0.008$ for 
$e\nu ee$, $\mu\nu ee$, $e\nu\mu\mu$ and $\mu\nu\mu\mu$ final states,
respectively.

Lepton-identification and central-track-matching efficiencies are
estimated using samples of $Z\rightarrow e^{+}e^{-}$ and 
$Z\rightarrow\mu^{+}\mu^{-}$ events.
One of the leptons from the $Z$ boson decay is required to pass all
lepton selection requirements. The other lepton is tested as to
whether it passes the selection criteria.
Both identification efficiencies and track-matching efficiencies are
determined as functions of $p_{T}$ and $\eta$. Average identification 
efficiencies are $0.929\pm0.013$ and $0.965\pm0.008$ for CC and EC
electrons, respectively, and $0.940\pm0.002$ for muons. 
Track-matching efficiencies are $0.817\pm0.002$ 
for CC electrons, $0.674\pm0.006$ for EC electrons, and $0.950\pm0.002$
for muons.  These efficiencies are folded into the {\it WZ} MC events used for
acceptance calculations. The overall {\it WZ} acceptance 
times detection
efficiencies are (10.3$\pm$1.5)\%, (11.7$\pm$0.8)\%, (13.9$\pm$1.3)\%,
and (16.3$\pm$1.8)\% for $e\nu ee$, $\mu\nu ee$, $e\nu \mu\mu$ and 
$\mu\nu\mu\mu$, respectively.

From the SM prediction for $\sigma_{\it WZ}$ and the leptonic branching
fractions of the $W$ and $Z$ bosons~\cite{ref:epj}, we expect
0.44$\pm$0.07, 0.45$\pm$0.04, 0.53$\pm$0.06, 0.62$\pm$0.08 {\it WZ} events for 
the $e\nu ee$, $\mu\nu ee$, $e\nu \mu\mu$, and $\mu\nu\mu\mu$ final 
states, respectively. 
Quoted uncertainties include statistical and systematic contributions, 
as well as the 6.5\% uncertainty in the integrated luminosity.

Among SM processes, {\it WZ} production is the dominant 
mechanism that results in events with a  
final state that includes three isolated leptons with large transverse 
momentum and with 
large \mbox{${\hbox{$E$\kern-0.6em\lower-.1ex\hbox{/}}}_T$}.
The main backgrounds to {\it WZ} production come from 
$Z+X$ ($X$=hadronic jets, 
$\gamma$, or $Z$) events. In $Z +{\rm jet(s)}$ events, a jet may be 
misidentified as an additional lepton. This background is estimated from 
data as follows. Events are selected using the same criteria as for the 
{\it WZ} sample, except that the requirement of the third lepton is 
dropped. The resulting ``dilepton + jet(s)'' sample includes $ee+{\rm jets}$, 
$\mu\mu+{\rm jets}$ and $e\mu+{\rm jets}$ events. Probabilities for 
hadronic jets to mimic electrons and muons are determined, 
using multi-jet data, as a function of jet $E_{T}$ and jet $\eta$. 
Applying the misidentification probabilities to jets in the dilepton + jet(s) 
events yields the total background, estimated to be $0.35\pm0.02$ events. 
In $Z+\gamma$ events, a $\gamma $ may be converted to electrons or randomly 
match a charged-particle track in the detector causing it to be misidentified 
as an electron. This background process only contributes to the 
$e\nu\mu\mu$ and $e\nu ee$ final states. Though we have identified hundreds
of $Z+\gamma$ events~\cite{r2zg}, we found the probability for a 
photon to be misidentified as an electron is $\sim 2\%$. As these 
events do not typically have large 
\mbox{${\hbox{$E$\kern-0.6em\lower-.1ex\hbox{/}}}_T$}, the number which
mimic the {\it WZ} signal is small. We estimate it as $0.145\pm0.020$ events.
The backgrounds from {\it ZZ} and $t\bar{t}$ production 
are estimated using Monte Carlo methods to be $0.20\pm0.07$ and $0.01\pm0.01$ 
events, respectively. Other sources of background are found
to be negligible. The total background is estimated to be 
$0.71\pm0.08$ events.

The combination of expected {\it WZ} signal and 
background is consistent with having observed three {\it WZ} candidates.
The probability for a background of $0.71$ events alone to fluctuate to three
or more candidates is 3.5\%.
Following the method described in Refs.~\cite{ref:epj} 
and \cite{ref:signf}, we use a maximum likelihood technique to obtain 
$\sigma_{\it WZ}=4.5_{-2.6}^{+3.8}$ $\mathrm{pb}$ and calculate the 95\% C.L.
upper limit $\sigma _{\it WZ}<$ $13.3$~\textrm{pb }for $\sqrt{s}=1.96$ 
\textrm{TeV}.

\begin{figure}[tbh]
\hskip 0cm \includegraphics[width=3.1 in]{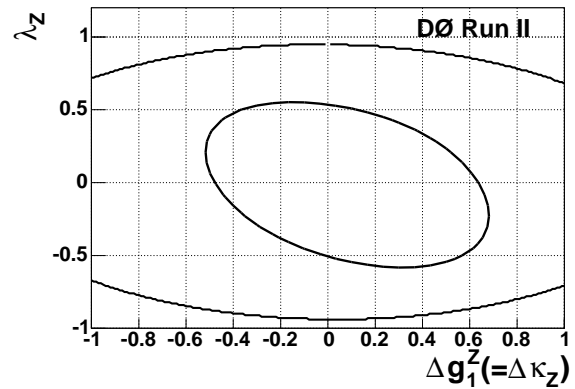}
\caption{Two-dimensional coupling limits (inner contour) 
on $\lambda_Z$ vs. $\Delta g_{1}^{Z}$ at 95\% C.L. for $\Lambda=1.5$ TeV 
under the assumptions of Ref.~\cite{LEPass},
which reduce to $\Delta \kappa_Z = \Delta g_{1}^{Z}$ for {\it WZ} production.
The outer contour is the limit from $S$-matrix unitarity.} 
\label{fig:countour}
\end{figure}

As $\sigma_{\it WZ}$ is consistent with the SM, we can  extract 
limits on anomalous {\it WWZ} couplings.  
Monte Carlo $WZ\rightarrow \,$trilepton events are 
generated~\cite{ref:wzgenerator} 
at each point in a two-dimensional grid of anomalous couplings. 
We used a parameterized detector simulation to model the detector response 
and applied the same selection criteria that were applied to the data to 
determine the predicted {\it WZ} signal at each grid point. These
predictions are combined with the estimated background and compared with 
the three observed trilepton candidates to construct a likelihood function 
$L$. Analyses of contours of $L$  then permits
limits to be set on $\lambda_{Z}$, $\Delta g_{1}^{Z}$ and $\Delta\kappa_{Z}$, 
both individually and in pairs, where  
$\Delta \kappa_Z \equiv \kappa_Z -1$ and 
$\Delta g_1^Z \equiv g_1^Z-1$.
Table~\ref{tab:limits} lists one-dimensional 95\%
C.L. limits on $\lambda_{Z}$, $\Delta g_{1}^{Z}$ and $\Delta\kappa_{Z}$ with 
$\Lambda=$1 TeV or $\Lambda=$1.5 TeV.  Figure~\ref{fig:countour} shows
two-dimensional 95\% C.L. contour limits for $\Lambda$ = 1.5 TeV with 
the assumption of $SU(2)_L \otimes U(1)_Y$ gauge invariance relating 
the couplings~\cite{LEPass}.  The values
of the form factors are chosen such that the coupling limit contours are
within the contours provided by $S$-matrix unitarity~\cite{ref:unitarity}.

\begin{table*}
\caption{One-dimensional 95\% C.L. intervals on $\protect\lambda_{Z}$, 
$\Delta g_{1}^{Z}$, and $\Delta\protect\kappa_{Z}$.
 In the missing last entry, 
the 95\% C.L. limit exceeded the bounds from $S$-matrix unitarity.
The assumption $\Delta g_1^Z  = \Delta \kappa_Z$ is equivalent to 
that used in Ref.~\cite{LEPass}.}
\begin{center}
\begin{ruledtabular}
\begin{tabular}{ccc}
Condition          & $\Lambda$ = 1 TeV   & $\Lambda$ = 1.5 TeV           \\ \hline
$\Delta g^Z_1=\Delta \kappa_Z=0$
                   & $-0.53 < \lambda_Z < 0.56$  
                                         & $-0.48 < \lambda_Z < 0.48$ \\
$\lambda_Z=\Delta \kappa_Z=0$ 
                   & $-0.57 < \Delta g^Z_1 < 0.76$
                                         & $-0.49 < \Delta g^Z_1 < 0.66$ \\
$\lambda_Z = 0$  
                   & $-0.49< \Delta g_1^Z  = \Delta \kappa_Z < 0.66$ 
                                         & $-0.43 < \Delta g_1^Z  
                                           = \Delta \kappa_Z < 0.57$  \\
$\lambda_Z=\Delta g^Z_1 =0 $
                   & $-2.0 < \Delta \kappa_Z < 2.4$   
                                         & $-$                         \\
\end{tabular}
\end{ruledtabular}
\end{center}
\label{tab:limits}
\end{table*}

In summary, we searched for {\it WZ} production in $p\bar{p}$ collisions at 
$\sqrt{s}=1.96$ TeV. In a sample of 0.30 fb$^{-1}$, 
three candidate events
were found with an expected background of $0.71\pm0.08$ events. The 95\% C.L. 
upper limit for the {\it WZ} cross section is $13.3$ pb. Interpreting the
candidates as a combination of {\it WZ} signal plus background,
we find $\sigma_{\it WZ}=4.5^{+3.8}_{-2.6}$ pb and
provide the first measurement of the {\it WZ} production cross
section at hadron colliders. We used the results of the search to
obtain the tightest available limits on anomalous {\it WWZ} couplings
derived from a {\it WZ} final state. Furthermore, these are the most
restrictive model-independent {\it WWZ} anomalous coupling limits 
available and represent an improvement by a factor of three over the previous 
best results~\cite{run1wz}. 


%
We thank the staffs at Fermilab and collaborating institutions, 
and acknowledge support from the 
DOE and NSF (USA),
CEA and CNRS/IN2P3 (France),
FASI, Rosatom and RFBR (Russia),
CAPES, CNPq, FAPERJ, FAPESP and FUNDUNESP (Brazil),
DAE and DST (India),
Colciencias (Colombia),
CONACyT (Mexico),
KRF (Korea),
CONICET and UBACyT (Argentina),
FOM (The Netherlands),
PPARC (United Kingdom),
MSMT (Czech Republic),
CRC Program, CFI, NSERC and WestGrid Project (Canada),
BMBF and DFG (Germany),
SFI (Ireland),
A.P.~Sloan Foundation,
Research Corporation,
Texas Advanced Research Program,
Alexander von Humboldt Foundation,
and the Marie Curie Fellowships.


\begin{thebibliography}{99}
\bibitem{coords} We use a right-handed coordinate system with $\hat{z}$ 
pointing in 
the direction of the proton beam and $\hat{y}$
pointing upwards.
\end{thebibliography}
\end{document}